\documentclass[12pt]{article}

\usepackage{latexsym,amsmath,amssymb,amsthm,amsfonts,graphicx,natbib}
\usepackage{epsfig}
\usepackage{bm}
\newcommand*\patchAmsMathEnvironmentForLineno[1]{%
      \expandafter\let\csname old#1\expandafter\endcsname\csname #1\endcsname
      \expandafter\let\csname oldend#1\expandafter\endcsname\csname end#1\endcsname
      \renewenvironment{#1}%
         {\linenomath\csname old#1\endcsname}%
         {\csname oldend#1\endcsname\endlinenomath}}%
    \newcommand*\patchBothAmsMathEnvironmentsForLineno[1]{%
      \patchAmsMathEnvironmentForLineno{#1}%
      \patchAmsMathEnvironmentForLineno{#1*}}%
    \AtBeginDocument{%
    \patchBothAmsMathEnvironmentsForLineno{equation}%
    \patchBothAmsMathEnvironmentsForLineno{align}%
    \patchBothAmsMathEnvironmentsForLineno{flalign}%
    \patchBothAmsMathEnvironmentsForLineno{alignat}%
    \patchBothAmsMathEnvironmentsForLineno{gather}%
    \patchBothAmsMathEnvironmentsForLineno{multline}%
    }
\usepackage[mathlines,displaymath]{lineno}
\include{def}
\def\dispmuskip{\thinmuskip= 3mu plus 0mu minus 2mu \medmuskip=  4mu plus 2mu minus 2mu \thickmuskip=5mu plus 5mu minus 2mu}
\def\textmuskip{\thinmuskip= 0mu                    \medmuskip=  1mu plus 1mu minus 1mu \thickmuskip=2mu plus 3mu minus 1mu}
\def\beq{\dispmuskip\begin{equation}}    \def\eeq{\end{equation}\textmuskip}
\def\beqn{\dispmuskip\begin{displaymath}}\def\eeqn{\end{displaymath}\textmuskip}
\def\bea{\dispmuskip\begin{eqnarray}}    \def\eea{\end{eqnarray}\textmuskip}
\def\bean{\dispmuskip\begin{eqnarray*}}  \def\eean{\end{eqnarray*}\textmuskip}

\def\paradot#1{\vspace{1.3ex plus 0.7ex minus 0.5ex}\noindent{\bf\boldmath{#1.}}}

\newcommand{\wh}{\widehat}
\newcommand{\wt}{\widetilde}

\def\E{{\rm E}}                         

\def\a{\alpha}

\def\s{\sigma}

\def\t{\theta}
\def\b{\beta}
\def\l{\lambda}

\def\M{{\cal M}}

\def\LPDS{\text{\rm LPDS}}

\def\Kl{\text{\rm KL}}

\def\MSE{\text{\rm MSE}}

\begin{document}

\title{Efficient variational inference for generalized \\
linear mixed models with large datasets}
\author{\normalsize David J Nott, Minh-Ngoc Tran,
Anthony Y.C. Kuk and Robert Kohn\footnote{  
David J. Nott is Associate Professor, Department of Statistics and Applied Probability,
National University of Singapore, Singapore 117546 (standj@nus.edu.sg). Minh-Ngoc Tran is Research Fellow,
Australian School of Business, University of New South Wales, Sydney 2052 Australia
(minh-ngoc.tran@unsw.edu.au).  Anthony Y.C. Kuk is Professor,  
Department of Statistics and Applied Probability, National University
of Singapore, Singapore 117546 (stakuka@nus.edu.sg).    
Robert Kohn is Professor, Australian School of Business, 
University of New South Wales, Sydney 2052 Australia (r.kohn@unsw.edu.au).}}
\maketitle
\begin{abstract}
The article develops a hybrid Variational Bayes algorithm that combines the mean-field and fixed-form Variational Bayes methods.
The new estimation algorithm can be used to approximate any posterior without relying on conjugate priors.
We propose a divide and recombine strategy for the analysis of large datasets,
which partitions a large dataset into smaller pieces and then combines the variational distributions
that have been learnt in parallel on each separate piece using the hybrid Variational Bayes algorithm. 
We also describe an efficient model selection strategy using cross validation,
which is trivial to implement as a by-product of the parallel run.
The proposed method is applied to fitting generalized linear mixed models.
The computational efficiency of the parallel and hybrid Variational Bayes algorithm
is demonstrated on several simulated and real datasets.

\paradot{Keywords} Parallelization, Mean-field Variational Bayes, Fixed-form Variational Bayes.

\end{abstract}
\section{Introduction}\label{sec:Introduction}
Variational Bayes (VB) methods are increasingly used in machine learning and statistics
as a computationally efficient alternative to Markov Chain Monte Carlo simulation
for approximating posterior distributions in Bayesian inference.
See, for example, \cite{Bishop:2006,Ormerod:2010}.
VB algorithms can be categorized into two main groups: the mean-field Variational Bayes (MFVB) algorithm 
\citep{Attias:1999,Waterhouse:1996,Ghahramani:2001} and the fixed-form Variational Bayes (FFVB) algorithm \citep{Honkela:2010,Salimans:2013}.
The MFVB algorithm provides an efficient and convenient iterative scheme 
for updating the variational parameters,
but in its exact form it requires conjugate priors and therefore rules out some interesting models.
The FFVB algorithm assumes a fixed functional form for the variational distribution
and employs some optimization approaches such as stochastic gradient descent search for estimating the variational parameters. 
This article develops a VB algorithm that combines these two algorithms
in which the stochastic search FFVB method of \cite{Salimans:2013} (see Section \ref{sec:Salimans and Knowles})
 is used within a MFVB procedure for updating variational distribution factors 
that do not have a conjugate form.
The convergence of the whole updating procedure is formally justified.
Related work by \cite{Waterhouse:1996} and \cite{Wang:2013} used the Laplace approximation 
for updating non-conjugate variational factors, and \cite{Knowles:2011} introduced the non-conjugate variational
message passing framework for variational Bayes with approximations in the exponential family when
the lower bound can be approxmiated in some way.  
 \cite{Braun:2010} and \cite{Wang:2013} also consider approximating
the lower bound in non-conjugate models using the delta method.  
\cite{Tan:2013b} extend the stochastic variational inference
approach of \cite{Hoffman:2013} by combining non-conjugate variational message passing with algorithms from stochastic optimization which work with mini-batches of data, 
and apply the idea to non-conjugate generalized linear mixed models. 
We refer to the suggested algorithm we develop as the fixed-form within mean-field Variational Bayes algorithm,
or the hybrid Variational Bayes algorithm.
The new algorithm can be used to conveniently and efficiently approximate any posterior without relying on the conjugacy assumption. 

The second contribution of this article is to propose a divide and recombine strategy \citep{Guha:2012}
for the analysis of large datasets based on exponential family variational Bayes posterior approximations.
The idea is to partition a large dataset into smaller pieces and learn 
the variational distribution in parallel on each separate piece using the hybrid Variational Bayes algorithm. 
The resulting variational distributions then are recombined to construct the final approximation of the posterior.
The recombination is particularly easy for posterior approximations in the exponential family.  
The methodology proposed in our article is closely related to the methodology proposed independently in a recent preprint by \cite{Broderick:2013}.
The main difference is that they develop the methodology in an online setting in which the data pieces arrive sequentially in time,
while we describe the method in a static setting in which the whole dataset has already been collected.
Furthermore, we show how to use the parallel divide and recombine strategy for model selection using cross validation.
We also study empirically the effect of the number of data pieces 
and recommend a good number to use in practice.

As a main application of the parallel and hybrid Variational Bayes algorithm,  
we derive a detailed algorithm for fitting generalized linear mixed models (GLMMs).
GLMMs are often considered difficult to estimate because of the presence of random effects and lack of conjugate priors.
VB schemes for GLMMs are considered previously by \cite{Rijmen:2008,Ormerod:2012} and \cite{Tan:2013},
and are shown to have attractive computation and accuracy trade-offs.
The computational efficiency and accuracy of the proposed method is demonstrated on several simulated and real datasets.

The rest of the paper is organized as follows.
Section \ref{sec:VB theory} provides the background to VB methods
and presents the hybrid VB algorithm.
Section \ref{sec:Salimans and Knowles} reviews the fixed-form VB method of \cite{Salimans:2013}
that we use for updating the non-conjugate variational factors within the mean-field VB algorithm.
Section \ref{sec:parallel} presents the parallel implementation idea for handling large datasets.
The detailed parallel and hybrid Variational Bayes algorithm 
for fitting GLMMs is presented in Section \ref{sec:application to GLMM},
and Section \ref{sec:examples} reports a simulation study and real data examples.

\section{Some Variational Bayes theory}\label{sec:VB theory}
Let $\theta$ be a vector of parameters, $p(\t)$ the prior and $y$ the data.
Variational Bayes (VB) approximates the posterior $p(\t|y)\propto p(\t)p(y|\t)$ by a more easily accessible distribution $q(\t)$,
which minimizes the Kullback-Leibler divergence
\beq\label{eq:KL_distance}
\Kl(q\|p) = \int q(\t)\log\frac{q(\t)}{p(\t|y)}d\t.
\eeq
We have 
\bean
\log p(y)&=&\int q(\t)\log\frac{p(y,\t)}{q(\t)}d\t+\int q(\t)\log\frac{q(\t)}{p(\t|y)}d\t\\
&=&L(q)+\Kl(q\|p),
\eean
where
\beq\label{eq:original VB}
L(q) = \int q(\t)\log\frac{p(y,\t)}{q(\t)}d\t.
\eeq
As $\Kl(q\|p)\geq0$, $\log\;p(y)\geq L(q)$ for every $q(\t)$, $L(q)$ is therefore often called the lower bound,
and minimizing $\Kl(q\|p)$ is therefore equivalent to maximizing $L(q)$.

Often factorized approximations to the posterior are considered in variational Bayes.  We explain the idea for a factorization with 2 blocks.
Assume that $\t=(\t_1,\t_2)$ and $q(\t)$ is factorized as 
\beq\label{eq:VBfull}
q(\t)=q(\t_1)q(\t_2).
\eeq
We further assume that $q(\t_1)=q_{\l_1}(\t_1)$ and $q(\t_2)=q_{\l_2}(\t_2)$ where $\l_1$ and $\l_2$ are variational parameters that need to be estimated.
Then
\bean
L(\l_1,\l_2)=L(q)&=&\int q_{\l_1}(\t_1)q_{\l_2}(\t_2)\log p(y,\t)d\t_1d\t_2-\int q_{\l_1}(\t_1)\log q_{\l_1}(\t_1)d\t_1+C(\l_2)\\
&=&\int q_{\l_1}(\t_1)\left(\int q_{\l_2}(\t_2)\log p(y,\t)d\t_2\right)d\t_1-\int q_{\l_1}(\t_1)\log q_{\l_1}(\t_1)d\t_1+C(\l_2)\\
&=&\int q_{\l_1}(\t_1)\log\wt p(y,\t_1)d\t_1-\int q_{\l_1}(\t_1)\log q_{\l_1}(\t_1)d\t_1+C(\l_2)\\
&=&\int q_{\l_1}(\t_1)\log\frac{\wt p_1(y,\t_1)}{q_{\l_1}(\t_1)}d\t_1+C(\l_2),
\eean
where $C(\l_2)$ is a constant depending only on $\l_2$ and 
\beqn
\wt p_1(y,\t_1)=\exp\left(\int q_{\l_2}(\t_2)\log p(y,\t)d\t_2\right)=\exp\big(\E_{-\t_1}(\log p(y,\t))\big).
\eeqn
Let
\beq\label{eq:lam1}
\l_1^* = \l_1^*(\l_2) = \arg\max_{\l_1}\left\{\int q_{\l_1}(\t_1)\log\frac{\wt p_1(y,\t_1)}{q_{\l_1}(\t_1)}d\t_1\right\},
\eeq
then 
\beq
L(\l_1^*,\l_2)\geq L(\l_1,\l_2)\;\;\text{for all}\; \l_1.
\eeq
Similarly, let
\beq\label{eq:lam2}
\l_2^* = \l_2^*(\l_1) = \arg\max_{\l_2}\left\{\int q_{\l_2}(\t_2)\log\frac{\wt p_2(y,\t_2)}{q_{\l_2}(\t_2)}d\t_2\right\},
\eeq
with
\beqn
\wt p_2(y,\t_2)=\exp\left(\int q_{\l_1}(\t_1)\log p(y,\t)d\t_1\right)=\exp\big(\E_{-\t_2}(\log p(y,\t))\big),
\eeqn
hence
\beq
L(\l_1,\l_2^*)\geq L(\l_1,\l_2)\;\;\text{for all}\; \l_2.
\eeq
Let $\l^\text{old}=(\l^\text{old}_1,\l^\text{old}_2)$ and $\l^\text{new}_1=\l_1^*(\l^\text{old}_2)$ as in \eqref{eq:lam1}
and $\l^\text{new}_2=\l_2^*(\l^\text{new}_1)$ in \eqref{eq:lam2}, we have 
\beq\label{eq:lb_increase}
L(\l^\text{new})\geq L(\l^\text{old}).
\eeq
This leads to an iterative scheme for updating $\l$
and \eqref{eq:lb_increase} ensures the improvement of the lower bound over the iterations.
Because the lower bound $L(\l)$ is bounded from above, the convergence of the iterative scheme is ensured under some mild conditions.
The above argument can be easily extended to the general case in which $q(\theta)$ is factorized into $K$ blocks $q(\t)=q(\t_1)\times...\times q(\t_K)$.

Variational Bayes approximation is now reduced to solving an optimization problem in form of \eqref{eq:lam1}.
In many cases, a conjugate prior $p(\t_1)$ can be selected such that $\wt p_1(y,\t_1)$ belongs to a recognizable density family,
then the optimal VB posterior $q_{\l_1}(\t_1)$ that maximizes the integral on the right hand side of \eqref{eq:lam1} is $\wt p_1(y,\t_1)$,
i.e.
\beq\label{eq:optimal_VB}
q_{\l_1^*}(\t_1)\propto \wt p_1(y,\t_1) = \exp\big(\E_{-\t_1}(\log p(y,\t))\big),
\eeq
and $\l_1^*$ is determined accordingly.
In such cases, the resulting iterative procedure is often referred to as the mean-field Variational Bayes (MFVB) algorithm,
or the Variational Bayes EM-like algorithm.
The MFVB is computationally convenient but it is not applicable to some interesting models
because of the requirement of conjugate priors.

If $\wt p_1(y,\t_1)$ does not belong to a recognizable density family,
some optimization technique is needed to solve \eqref{eq:lam1}.
Note that \eqref{eq:lam1} has exactly the same form as the original VB problem that attempts to maximize $L(q)$ in \eqref{eq:original VB}.
We can first select a functional form for the variational distribution $q$ and then estimate the 
unknown parameters accordingly. Such a method is known in the literature as the fixed-form Variational Bayes (FFVB) algorithm. 
If the variational distribution is assumed to belong to the exponential family with unknown parameters $\l$,
\cite{Salimans:2013} propose a stochastic approximation method for solving for $\l$.
The details of this method are presented in next section.
It is obvious that we can use a FFVB algorithm within a MFVB procedure to solve for \eqref{eq:lam1}
and the convergence of the whole procedure is still guaranteed. 
Interestingly, this procedure is similar in spirit to the popular Metropolis-Hastings within Gibbs sampling in Markov Chain Monte Carlo simulation. 
 
\section{Fixed-form Variational Bayes method of Salimans and Knowles}\label{sec:Salimans and Knowles}
Suppose we have data $y$, a likelihood $p(y|\theta)$ where $\theta\in\mathbb{R}^d$ is an unknown parameter, 
and a prior distribution $p(\theta)$ for $\theta$. 
\cite{Salimans:2013} approximate the posterior $p(\theta|y)\propto p(\theta)p(y|\theta)$ 
by a density (with respect to some base measure which for simplicity we assume is the Lebesgue measure below) which is in the exponential family
\begin{eqnarray*}
q_\lambda(\theta) & = & \exp\left( S(\theta)^T \lambda-Z(\lambda)\right), 
\end{eqnarray*}
where $\lambda$ is a vector of natural parameters, $S(\theta)$ denotes
a vector of sufficient statistics for the given exponential family and $Z(\lambda)$ is a normalization term.  
The $\l$ is chosen by minimizing the Kullback-Leibler divergence
\begin{eqnarray*}
  \Kl(\lambda)  =  \int \log \frac{p(\theta|y)}{q_\lambda(\theta)} q_\lambda(\theta) d\theta  = \int \left\{\log p(\theta|y)-S(\theta)^T \lambda + Z(\lambda)\right\} \exp\left(S(\theta)^T \lambda - Z(\lambda)\right)d\t.
\end{eqnarray*}
Differentiating with respect to $\lambda$, and using the result for exponential families that
\begin{eqnarray} 
 \nabla_\lambda Z(\lambda)  = \int S(\theta) q_\lambda(\theta) d\theta  = E_\lambda (S(\theta)), \label{effact}
\end{eqnarray}
which can be obtained by differentiating the normalization condition $\int q_\lambda(\theta)d\theta=1$ 
with respect to $\lambda$, we have
\begin{eqnarray*}
\nabla_\lambda \Kl(\lambda) & = & \int \left\{-S(\theta)+\nabla_\lambda Z(\lambda)\right\} q_\lambda(\theta)d\theta \\
 & & +\int \left\{S(\theta)-\nabla_\lambda Z(\lambda)\right\} \left\{\log p(\theta|y)-S(\theta)^T \lambda+Z(\lambda)\right\} q_\lambda(\theta)d\theta. 
\end{eqnarray*}
Using (\ref{effact}), the first term on the right hand side above disappears leaving
\begin{eqnarray*}
\nabla_\lambda \Kl(\lambda) & = & \int \log p(\theta|y)\left\{S(\theta)-\nabla_\lambda Z(\lambda)\right\} q_\lambda(\theta) d\theta \\
 & & -\int \left\{S(\theta) S(\theta)^T \lambda - \nabla_\lambda Z(\lambda) S(\theta)^T \lambda-S(\theta) Z(\lambda)+\nabla_\lambda Z(\lambda)Z(\lambda)\right\} q_\lambda(\theta) d\theta \\
 & = & \mbox{Cov}_\lambda \big(S(\theta),\log p(\theta|y)\big)-\mbox{Cov}_\lambda\big(S(\theta)\big)\lambda
\end{eqnarray*}
where in obtaining the last line we have again made use of (\ref{effact}).  Hence $\nabla_\lambda \Kl(\lambda)=0$ if
\begin{eqnarray}
 \lambda & = & \mbox{Cov}_\lambda\big(S(\theta)\big)^{-1} \mbox{Cov}_\lambda\big(S(\theta),\log p(\theta|y)\big). \label{opt1}
\end{eqnarray}
This is a fixed point iteration that holds for the optimal value of $\lambda$.  Note that $\log p(\theta|y)$ differs
only by a constant not depending on $\theta$ from $\log p(\theta)p(y|\theta)$ so (\ref{opt1}) can be written
\begin{eqnarray}
 \lambda & = & \mbox{Cov}_\lambda\big(S(\theta)\big)^{-1} \mbox{Cov}_\lambda\big(S(\theta),\log p(\theta)p(y|\theta)\big). \label{optcondn}
\end{eqnarray}
For minimization of $\Kl(\lambda)$ this suggests an iterative scheme where at iteration $k$ the parameters
$\lambda^{(k)}$ are updated to 
\begin{eqnarray}
\lambda^{(k+1)} & = & \mbox{Cov}_{\lambda^{(k)}}\big(S(\theta)\big)^{-1} \mbox{Cov}_{\lambda^{(k)}}\big(S(\theta),\log p(\theta)p(y|\theta)\big). \label{nupdate}
\end{eqnarray}
\cite{Salimans:2013} observe that this iterative scheme doesn't necessarily converge.  Instead, inspired by 
stochastic gradient descent algorithms \citep{Robbins:1951} they choose to estimate
$\mbox{Cov}_\lambda(S(\theta))$ and $\mbox{Cov}_\lambda(S(\theta),\log p(\theta)p(y|\theta))$ by a weighted
average over iterates in a Monte Carlo approximation to a pre-conditioned gradient descent algorithm which
is guaranteed to converge if a certain step size parameter in their algorithm is small enough. 
They argue for Monte Carlo estimation of both 
$\mbox{Cov}_{\lambda^{(k)}}(S(\theta))$ and $\mbox{Cov}_{\lambda^{(k)}}(S(\theta),\log p(\theta)p(y|\theta))$ using
the same Monte Carlo samples.  This results in the approximation of the right hand side of (\ref{nupdate}) 
taking the form of a linear regression of the log target distribution on the sufficient statistics of
the approximating family.  The number of iterations $N$ for which their algorithm is run is decided upon
in advance, a constant step size of $c=1/\sqrt{N}$ is chosen for all iterations and averaging is over the
last $N/2$ iterations in forming the estimates of the covariance matrices to calculate an estimate of 
$\lambda$.  Theoretical support for these choices in the context of stochastic
gradient descent algorithms is given by \cite{Nemirovski:2009}.  See \cite{Salimans:2013}
for further discussion of why stochastic estimation of the covariance matrices rather than averaging over the parameters
$\lambda$ in a more conventional stochastic gradient algorithm is beneficial.

\cite{Salimans:2013} show using properties of the exponential family that
\begin{eqnarray}
 \mbox{Cov}_\lambda(S(\theta)) & = & \nabla_\lambda E_\lambda(S(\theta)) \label{cov1},
\end{eqnarray}
and
\begin{eqnarray}
 \mbox{Cov}_\lambda\big(S(\theta),\log p(\theta)p(y|\theta)\big) & = & \nabla_\lambda E_\lambda\big(\log p(\theta)p(y|\theta)\big), \label{cov2}
\end{eqnarray}
and then consider Monte Carlo approximations to the expectations on the right hand side of (\ref{cov1}) and of
(\ref{cov2}) based on a random draw $\theta^*\sim q_\lambda(\theta)$ where $\theta^*=f(\lambda,s)$ 
and $s$ is some random seed.  If $f$ is smooth, the Monte Carlo approximations are smooth functions of $\lambda$, 
and these approximations can be differentiated in (\ref{cov1}) and (\ref{cov2}).  In the case of 
an approximating distribution $q_\lambda(\theta)$ which is multivariate normal, and working in a direct
parameterization in terms of the mean and covariance matrix, results due to \cite{Minka:2001} and \cite{Opper:2009}
are used for evaluating the gradients in (\ref{cov1}) and (\ref{cov2}) to simplify the approximations while
making use of first and second derivative information of the target posterior \citep[][Section 4.4
and Appendix C]{Salimans:2013}.  This results in a highly efficient algorithm.  

We will be concerned with a certain modification of their algorithm for Gaussian $q_\lambda(\theta)$ but
where there is independence between blocks of the parameters.  Suppose $\theta$ is decomposed into $K$ blocks
$\theta=(\theta_1^T,...,\theta_K^T)^T$ and that the variational posterior $q_\lambda(\theta)$ factorizes
as
$$q_\lambda(\theta)=q_{\lambda_1}(\theta_1)\times ...\times q_{\lambda_K}(\theta_K)$$
with each factor $q_{\lambda_k}(\theta_k)$, $k=1,\dots ,K$, being multivariate normal.  Here 
$\lambda_k$ denotes the natural parameter for the $k$th factor, we write $\mu_k$ and $\Sigma_k$ for
the corresponding mean and covariance matrix and write $S_k(\theta_k)$ for the
vector of sufficient statistics in the $k$th normal factor.  
Because of the independence, the optimality condition
(\ref{optcondn}) simplifies to 
$$\lambda_k=\mbox{Cov}_{\lambda_k}\big(S_k(\theta_k)\big)^{-1} \mbox{Cov}_\lambda\big(S_k(\theta_k),\log p(\theta)p(y|\theta)\big)$$
and we can use the ideas of \cite{Salimans:2013} to estimate the covariance matrices
on the right hand side of this expression.  The result is the following slight modification of
their Algorithm 2.  In the description below $t_k$, $g_k$, $\bar{t}_k$, 
$\bar{g}_k$ are vectors of the same length as $\theta^{(j)}$ and $\Gamma_k$ and $\bar{\Gamma}_k$ are square
matrices with dimension the length of $\theta_k$.  We assume below that $N$ is even
so that $N/2$ is an integer and set $c=1/\sqrt{N}$.

\noindent 
{\bf Algorithm 1:} 
\begin{itemize}
\item Initialize $\mu_k,\Sigma_k$, $k=1,\dots,K$.  
\item Initialize $t_k=\mu_k$, $\Gamma_k=\Sigma_k^{-1}$ and $g_k=0$, $k=1,\dots ,K$.  
\item Initialize $\bar{t}_k=0$, $\bar{\Gamma}_k=0$ and $\bar{g}_k=0$, $k=1,\dots ,K$.  
\item For $i=1,\dots,N$ do
\begin{itemize}
\item Generate a draw $\theta^*=(\theta_1^*,...,\theta_K^*)^T$ from $q_\lambda(\theta)$
\item For $k=1,\dots ,K$ do
\begin{itemize}
\item Set $\Sigma_k=\Gamma_k^{-1}$ and $\mu_k=\Sigma_kg_k+t_k$
\item Calculate the gradient $g_i^{(k)}$ and Hessian $H_i^{(k)}$ of $\log p(\theta)p(y|\theta)$ with respect to $\theta_k$ evaluated at $\theta^*$.  
\item Set $g_k=(1-c)g_k+c g_i^{(k)}$, $\Gamma_k=(1-c)\Gamma_k-c H_i^{(k)}$, $t_k=(1-c)t_k+c\theta_k^*$.
\item If $i>N/2$ then set $\bar{g}_k=\bar{g}_k+\frac{2}{N}g_i^{(k)}$, $\bar{\Gamma}_k=\bar{\Gamma}_k-\frac{2}{N}H_i^{(k)}$, $\bar{t}_k=\bar{t}_k+\frac{2}{N}\theta_k^*$.
\end{itemize}
\end{itemize}
\item Set $\Sigma_k=\bar{\Gamma}_k^{-1}$, $\mu_k=\Sigma_k\bar{g}_k+\bar{t}_k$ for $k=1,\dots ,K$. 
\end{itemize}

On termination of the algorithm $\mu_k$, $\Sigma_k$ are the estimated mean and covariance matrix in the normal term
$q_{\lambda_k}(\theta_k)$.  

\section{Parallel implementation for large datasets}\label{sec:parallel}
Suppose the data $y$ are partitioned into $M$ pieces, $y'=({y^{(1)}}',...,{y^{(M)}}')'$.  Suppose also that we have
learnt a variational posterior distribution for each piece, 
$q_{\lambda^{(j)}}(\theta)$ approximating $p(\theta|y^{(j)})$.  We assume that
\beq  \label{fact}
q_{\lambda^{(j)}}(\theta)  =  q_{\lambda^{(j)}_1}(\theta_1)\times \dots \times q_{\lambda^{(j)}_K}(\theta_K)
\eeq
where $\lambda^{(j)}_k$ is the natural parameter for $q_{\lambda^{(j)}_k}(\theta_k)$ which has been assumed to
have an exponential family form, $j=1,...,M$ and $k=1,...,K$.  
We will also assume that
\beqn
p(y|\theta)=p(y^{(1)}|\theta)\times \dots \times p(y^{(M)}|\theta)
\eeqn
i.e. the blocks $y^{(1)},...,y^{(M)}$ are conditionally independent given $\theta$.  Then
the posterior distribution is 
\bean
  p(\theta|y) & \propto & p(\theta)p(y^{(1)}|\theta) \times \dots \times p(y^{(M)}|\theta) \\
 & = & \frac{\left\{p(\theta)p(y^{(1)}|\theta)\right\}\times \dots \times \left\{p(\theta)p(y^{(M)}|\theta)\right\}}
 {p(\theta)^{M-1}} \\
 & \propto & \frac{p(\theta|y^{(1)})\times \dots \times p(\theta|y^{(M)})}{p(\theta)^{M-1}}.
\eean
Hence given our approximation $q_{\lambda^{(j)}}(\theta)$ of $p(\theta|y^{(j)})$, $p(\theta|y)$ is approximately
proportional to 
\beqn
\frac{q_{\lambda^{(1)}}(\theta) \times \dots \times q_{\lambda^{(M)}}(\theta)}{p(\theta)^{M-1}}.
\eeqn
The reasoning used here is the same as that used in the Bayesian committee machine \citep{Tresp:2000}
although Tresp focused more on applications to Gaussian process regression. 
A similar strategy was independently proposed in a recent preprint by \cite{Broderick:2013},
who assume that the data pieces $y^{(j)}$ arrive sequentially in time.
 
Recall that $q_{\lambda^{(j)}}(\theta)$ has the factorization (\ref{fact}) so that if
the prior also factorizes
\beqn
p(\theta)=p_{\lambda^{(0)}_1}(\theta_1)\times \dots \times p_{\lambda^{(0)}_K}(\theta_K)
\eeqn
where $p_{\lambda^{(0)}_k}(\theta_k)$, with natural parameters $\lambda^{(0)}_k$, has the same exponential family form as 
$q_{\lambda^{(j)}_k}(\theta_k)$, 
then the marginal posterior for $\theta_k$ is approximately proportional to
\beq\label{e:recombine}
\frac{q_{\lambda^{(1)}_k}(\theta_k)\times \dots \times q_{\lambda^{(M)}_k}(\theta_k)}{p_{\lambda^{(0)}_k}^{M-1}(\theta_k)}, k=1,...,K.
\eeq
This approximation to $p(\theta_k|y)$ is an exponential family distribution of the
same form as each of the factors with natural parameter $\sum_{j=1}^M \lambda^{(j)}_k-(M-1)\lambda^{(0)}_k$.  Hence
we can learn the approximations $q_{\lambda^{(j)}}(\theta)$ independently in parallel for different
chunks of the data and then combine these posteriors to get an approximation to the full posterior.

If the factors $q_{\lambda^{(j)}_k}(\theta_k)$ are all normal, with $\lambda^{(j)}_k$ corresponding
to mean $\mu^{(j)}_k$ and covariance matrix $\Sigma^{(j)}_k$ and if $p_{\lambda^{(0)}_k}(\theta_k)$ 
has mean $\mu^{(0)}_k$ and covariance matrix $\Sigma^{(0)}_k$, then the approximation to 
$p(\theta_k|y)$ is normal, with mean
\beqn
\left(\sum_{j=1}^M {\Sigma^{(j)}_k}^{-1}-(M-1){\Sigma^{(0)}_k}^{-1}\right)^{-1}\left(\sum_{j=1}^M {\Sigma^{(j)}_k}^{-1}\mu^{(j)}_k
-(M-1){\Sigma^{(0)}_k}^{-1}\mu^{(0)}_k\right)
\eeqn
and covariance matrix
\beqn
\left(\sum_{j=1}^M {\Sigma^{(j)}_k}^{-1}-(M-1){\Sigma^{(0)}_k}^{-1}\right)^{-1}.
\eeqn
A similar way of combining normal approximations of posterior distributions in mixed
models has been considered by Huang and Gelman (2005).  
If $q_{\lambda^{(j)}_k}(\theta_k)$ is Wishart, $W(\nu^{(j)}_k,S^{(j)}_k)$, 
and if $p_{\lambda^{(0)}_k}(\theta_k)$ is
Wishart, $W(\nu^{(0)}_k,S^{(0)}_k)$, then $p(\theta_k|y)$ is approximated as Wishart, 
\beqn
W\left(\sum_{j=1}^M \nu^{(j)}_k-(M-1)\nu^{(0)}_k,\left(\sum_{j=1}^M {S^{(j)}_k}^{-1}-(M-1) {S^{(0)}_k}^{-1}\right)^{-1}\right).
\eeqn  

\subsection{Model selection with cross-validation}\label{subsec:model selection}
The way of combining approximations learnt independently on different pieces of
the data makes model choice by cross-validation trivial to implement. 
Let one of the pieces $y^{(1)}$,...,$y^{(M)}$ be a future dataset $y_F$, 
and the rest is used as the training data $y_T$.     
Let $\M$ be the model that is being considered.
A common measure of the performance of the model $\M$ is the log predictive density scores (LPDS) defined as \citep{Good:1952}
\beqn
\LPDS(\M) = \log p(y_F|y_T,\M) = \log\int p(y_F|\theta,\M)p(\theta|y_T,\M)d\t,
\eeqn
where we assume that $p(y_F|\theta,y_T,\M)=p(y_F|\theta,\M)$, i.e. conditional on $\M$ and $\theta$ the future observations are independent of the observed,
and $p(\theta|y_T,\M)$ is the posterior of the model parameter $\theta$ conditional on the training data $y_T$.
The posterior $p(\theta|y_T,\M)$ can be replaced by its VB estimate $q(\theta|y_T,\M)$
and $\LPDS(\M)$ then can be approximated by Monte Carlo samples drawn from $q(\theta|y_T,\M)$.
A simpler method is to estimate the integral by $p(y_F|\wh\theta(y_T),\M)$
with $\wh\theta(y_T)$ an estimator of the posterior mean of $\t$ which can be obtained by using the mean of the VB approximation $q(\theta|y_T,\M)$.
We use this plug-in method in this paper and define 
the $M$-fold cross-validated LPDS as 
\beq\label{e:LPDS}
\LPDS(\M) = \frac{1}{M}\sum_{j=1}^M\log p(y^{(j)}|y\setminus y^{(j)},\M)\approx \frac{1}{M}\sum_{j=1}^M\log p(y^{(j)}|\wh\t(y\setminus y^{(j)}),\M).
\eeq
Computing \eqref{e:LPDS} is trivial with parallel implementation
and the main advantage is that no extra time is needed to refit the model on each training dataset. 
From \eqref{e:recombine}, the variational distribution $q(\t_k|y\setminus y^{(j)},\M)$ of the parameter block $\t_k$ conditional on dataset $y\setminus y^{(j)}$ is proportional to
\beqn
\frac{q_{\lambda^{(1)}_k}(\theta_k,\M)\times \dots \times q_{\lambda^{(j-1)}_k}(\theta_k,\M)\times q_{\lambda^{(j+1)}_k}(\theta_k,\M) \times\dots \times q_{\lambda^{(M)}_k}(\theta_k,\M)}{p_{\lambda^{(0)}_k}^{M-2}(\theta_k,\M)}, k=1,...,K,
\eeqn
from which the estimator $\wh\theta(y\setminus y^{(j)})$ is easily computed accordingly.
Recall that $q_{\lambda^{(j)}_k}(\theta_k,\M)$ is the VB approximation to the marginal posterior of the $k$th block $\t_k$, based on the $j$th data piece, $j=1,...,M$ and $k=1,...,K$.

\section{Application to generalized linear mixed models}\label{sec:application to GLMM}
Consider a generalized linear mixed model in which given random effects $b_i$ there are
vectors of responses $y_i=(y_{i1},...,y_{in_i})^T$, $i=1,...,m$, where the $y_{ij}$ are
conditionally independently distributed with a distribution in an exponential family
with density or probability function
$$f(y_{ij}|\beta,b_i)=\exp\left(\frac{y_{ij}\eta_{ij}-b(\eta_{ij})}{\phi}+c(y_{ij},\phi)\right)$$
where $\eta_{ij}$ is a canonical parameter which is monotonically related to the conditional
mean $\mu_{ij}=E(y_{ij}|\beta,b_i)$ through a link function $g(\cdot)$, $g(\mu_{ij})=\eta_{ij}$,
$\beta$ is a $p$-vector of fixed effect parameters, $\phi$ is a scale parameter which we assume
known (for example, in the binomial and Poisson families $\phi=1$), 
and $b(\cdot)$ and $c(\cdot)$ are known functions.  
Here for simplicity we are considering the case of a canonical link function,
i.e. $g(\mu_{ij})=\eta_{ij}$.  
The vector $\eta_i=(\eta_{i1},...,\eta_{in_i})^T$ is modeled as
$\eta_i=X_i\beta+Z_i b_i$,
where $X_i$ is an $n_i\times p$ design matrix for the
fixed effects and $Z_i$ is an $n_i\times u$ matrix of random effects (where $u$ is the dimension
of $b_i$).  
Let $b=(b_1',...,b_m')'$ and
\[
X = \begin{pmatrix}
X_1\\
X_2\\
\vdots\\
X_m
\end{pmatrix},\;\;Z=\begin{pmatrix}
Z_1&0&\cdots&0\\
0&Z_2&\cdots&0\\
\vdots&\vdots&\cdots&\vdots\\
0&0&\cdots&Z_m
\end{pmatrix},\;\;\eta = \begin{pmatrix}
\eta_1\\
\eta_2\\
\vdots\\
\eta_m
\end{pmatrix}=X\beta+Zb.
\]
The likelihood can be written as
$$p(y|\beta,b)=\prod_{i=1}^m\prod_{j=1}^{n_i}f(y_{ij}|\beta,b_i)=\exp\left(\frac{1}{\phi}(y^T\eta-1^Tb(\eta))+c(y,\phi)\right),$$
where $b(\eta)$ is understood componentwise and $c(y,\phi)=\sum_{i,j}c(y_{ij},\phi)$.
The random effects $b_i$ are independently distributed as $N(0,Q^{-1})$,
therefore $p(b)\sim N(0,Q^{-1}_b)$ with $Q_b$ a block diagonal matrix $\text{diag}(Q,...,Q)$.
 
We consider Bayesian inference, with a normal prior for $\beta$, $N(\mu_\beta^0,\Sigma_\beta^0)$ say, 
where $\mu_\beta^0$ and $\Sigma_\beta^0$ are known hyperparameters. 
The prior for $Q$ is a Wishart $W(\nu_0,S_0)$ where again $\nu_0$ and $S_0$ are known hyperparameters.
We set $\mu_\beta^0=0$, $\Sigma_\beta^0=\tau_0I_p$, $\nu_0=u+1$ and $S_0=\tau_0I_u$ with a large $\tau_0$, 1000 say.

We write $\theta=(\beta,b,Q)$ for the vector of all the unknown parameters and random effects,
and assume a factorized form for the variational posterior 
$$q(\theta)=q(\beta,b) q(Q)=q(\a) q(Q)\;\;\;\;\text{with}\;\;\;\;\alpha=(\beta^T,b^T)^T,$$
where $q(\a)$ is normal with mean $\mu_\a^q$ and covariance matrix $\Sigma_\a^q$, 
and $q(Q)$ is Wishart $W(\nu^q,S^q)$.  
It is important to note that treating $\b$ and $b$ as a single block rather than as two independent blocks
has a big influence on the statistical inferences because of strong posterior dependence between the fixed and random effects.

By combining the VB theory in Section \ref{sec:Introduction} and Algorithm 2 of Section \ref{sec:Salimans and Knowles}, 
we have the following mean-field fixed-form VB algorithm for fitting GLMM.

\noindent {\bf Algorithm 2}  
\begin{enumerate}
\item Initialize $\nu^q, S^q$.  
\item Update $\mu_\a^q$ and $\Sigma_\a^q$ as follows 
\begin{itemize}
\item Initialize $t_\a=\mu_\a^q$, $\Gamma_\a={\Sigma_\a^q}^{-1}$ and $g_\a=0$.  
\item Initialize $\bar{t}_\a=0$, $\bar{\Gamma}_\a=0$ and $\bar{g}_\a=0$.
\item For $i=1,\dots,N$ do
\begin{itemize}
\item Generate $\a^*=({\beta^*}^T,{b^*}^T)^T\sim N(\mu_\a^q,\Sigma_\a^q)$ and compute $\eta^*=X\beta^*+Zb^*$.  
\item Set $\Sigma_\a^q=\Gamma_\a^{-1}$ and $\mu_\a^q=\Sigma_\a^q g_\a+t_\a$.
\item Compute the gradient 
\beqn
g_i^\a = \begin{pmatrix}
\frac{1}{\phi}X^T(y-\dot{b}(\eta^*))-{\Sigma_{\a}^0}^{-1}(\beta^*-\mu_{\beta}^0)\\
\frac{1}{\phi}Z^T(y-\dot{b}(\eta^*))-\E_{q(Q)}(Q_b)b^*
\end{pmatrix}
\eeqn
and Hessian 
\beqn
H_i^\a = 
\begin{pmatrix}
-\frac{1}{\phi}X^T\mbox{diag}\left(\ddot{b}(\eta^*)\right)X - {\Sigma_{\a}^0}^{-1}& -\frac{1}{\phi}X^T\mbox{diag}\left(\ddot{b}(\eta^*)\right)Z \\
-\frac{1}{\phi}Z^T\mbox{diag}\left(\ddot{b}(\eta^*)\right)X& -\frac{1}{\phi}Z^T \mbox{diag}\left(\ddot{b}(\eta^*)\right)Z - \E_{q(Q)}(Q_b)
\end{pmatrix}
\eeqn
\item Set $g_\a=(1-c)g_\a+cg_i^\a$, $\Gamma_\a=(1-c)\Gamma_\a-cH_i^\a$, $t_\a=(1-c)t_\a+c\a^*$.  
\item If $i>N/2$ then set $\bar{g}_\a=\bar{g}_\a+\frac{2}{N}g_i^\a$, $\bar{\Gamma}_\a=\bar{\Gamma}_\a-\frac{2}{N}H_i^\a$, $\bar{t}_\a=\bar{t}_\a+\frac{2}{N}\a^*$,
\end{itemize}
\item Set $\Sigma_\a^q=\bar{\Gamma}_\a^{-1}$, $\mu_\a^q=\Sigma_\a^q\bar{g}_\a+\bar{t}_\a$.   
\end{itemize}
\item Update $\nu^q=\nu_0+m$, $S^q=\left(S_0^{-1}+\sum_{i=1}^m(\mu_{b_i}^q{\mu_{b_i}^q}^T+\Sigma_{b_i}^q)\right)^{-1}$.  
\item Repeat Steps 2-3 until convergence.
\end{enumerate}
In the above algorithm $\E_{q(Q)}(Q_b)=\text{diag}(\E_{q(Q)}(Q),...,\E_{q(Q)}(Q))$ with $\E_{q(Q)}(Q)=\nu^qS^q$,
$\mu_{b_i}^q$ and $\Sigma_{b_i}^q$ are mean and covariance of $b_i$ computed accordingly from $\mu_{\a}^q$ and $\Sigma_\a^q$. 
The $H_i^\a$ and therefore $\bar\Gamma_\a$ are block, high-dimensional and sparse matrices
whose lower right blocks are block diagonal.
Techniques of handling such sparse matrices should be used to reduce the computing time.     
Specially, we should compute the inverse of $\bar\Gamma_\a$ in blocks.
In our experience, the algorithm often converges very quickly, after a few iterations.
A common stopping rule is to stop iterations when the lower bound is not improved any further.
However, computing the lower bound in the GLMM context often involves an analytically intractable integral.
Alternatively, we can stop iterations if the difference of the variational parameters between two successive iterations is smaller than a small threshold.
In our implementation, the algorithm is terminated if either $1/d$ times the difference between two successive iterations is smaller than $\epsilon=10^{-5}$ ($d$ is the total number of the parameters)
or the number of iterations exceeds 50.
The number of iterations within each fixed-form update $N$ is set to 100 after some experimentation
but this can be varied depending on the computational budget or even adaptively increased as we near convergence.

In the GLMMs context, the data consist of observations on $m$ subjects $(y_i,X_i,Z_i),\ i=1,...,m$. 
To carry out the parallel implementation for large datasets,
we randomly partition the whole dataset of $m$ subjects into $M$ pieces such that each piece has $m_j\approx 200$ subjects (see Section \ref{subsec:simulations}),
$\sum_1^Mm_j=m$.
The variational distribution is learnt separately and in parallel on each piece using Algorithm 2,
and then recombined as in Section \ref{sec:parallel}.

\subsection{Model selection for GLMMs}\label{subsec:model selection GLMM}
Given the response vector $y$, assume that a GLMM has been specified,
then model selection in GLMMs consists of selecting fixed effect covariates and random effect covariates among a set of potential covariates.
Assume that we have fitted a GLMM $\M$ 
and denote the estimated parameter by $\wh\theta=(\wh\beta,\wh Q)$.
The log predictive score of a future dataset with response vector $y_F$, fixed effect design matrix $X_F$ and random effect design matrix $Z_F$ is
\beqn
\log p(y_F|\wh\t,\M) = \sum_{y_i\in y_F}\log\int \exp\left(\frac{1}{\phi}(y_i^T\eta_i-1^Tb(\eta_i))+c(y_i,\phi)\right)p(b_i|\wh Q)db_i,
\eeqn
where $\eta_i=X_i\wh\beta+Z_ib_i$ and $X_i\in X_F$, $Z_i\in Z_F$, correspondingly.
The integrals above can be estimated by the Laplace method.
The $M$-fold cross-validated LPDS is then computed as in \eqref{e:LPDS}.
The model that has the biggest LPDS will be selected.

It is obvious that this model selection strategy can be used for selecting GLMMs themselves as well as the link functions. 
A drawback of this model selection method is that it is not suitable for 
cases in which the number of candidate covariates is large because the total number of candidate models is huge 
and searching over the model space is very time demanding.  

\section{Examples}\label{sec:examples}
The proposed hybrid VB algorithm is written in Matlab and run on an Intel Core 16 i7 3.2GHz desktop.
The parallel implementation is supported by the Matlab Parallel Toolbox with 4 local workers.

The performance of the suggested VB method is compared to a MCMC simulation method.
If the likelihood is estimated unbiasedly, then the Metropolis-Hastings algorithm with the likelihood replaced by 
its unbiased estimator is still able to sample
exactly from the posterior. See, for example, \cite{Andrieu:2009} and \cite{Flury:2011}.
The likelihood in the GLMM context is a product of $m$ integrals over the random effects.
Each integral is estimated using importance sampling,  
which uses $S=10$ samples and the Laplace approximation for selecting the importance proposal density.
Note that each likelihood estimation is also run in parallel using the \texttt{parfor} loop in the Parallel Toolbox.
To handle the positive definiteness constraint on the inverse covariance $Q$,
we use the Leonard and Hsu transformation \citep{Leonard:1992} $Q = \exp(\Sigma)$, 
where $\Sigma$ is an unconstrained symmetric matrix,
to reparameterize $Q$ by the lower-triangle elements $\theta_Q$ of $\Sigma$, which is an one-to-one transformation between $Q$ and $\theta_Q$.
We then use the adaptive random walk Metropolis-Hastings algorithm in \cite{Haario:2001}
to sample from the posterior $p(\beta,\theta_Q|y)$.
Each MCMC chain consists of 20000 iterates with another 20000 iterates as burn-in.

Alternative MCMC methods for estimating GLMMs such as Gibbs sampling \citep{Zeger:1991}
can be faster than the MCMC scheme implemented in this paper.
However, the Metropolis-Hastings sampling scheme with the random effects integrated out using importance sampling can avoid 
mixing problems that one would have with Gibbs sampling due to strong coupling of the fixed and random effects.  
Gibbs sampling and similar MCMC methods for GLMMs are in general not parallelizable and therefore cumbersome in cases of a very large $m$.
It should be noted that it is often difficult to compare the CPU times between different algorithms
which depend heavily on the programming language being used and the optimality of the algorithms implemented
for the characteristics of the particular example considered.  However, we believe the results reported here are indicative
of the speed up obtained with our variational Bayes methods.

\subsection{Simulations}\label{subsec:simulations}
\subsubsection{A simulation study of parallel implementation}
This simulation example studies the effect of the divide and recombine strategy 
and its parallel implementation.
We consider the following logistic model with a random intercept
\bea\label{e:bino sil}
y_{ij}&\sim&\text{Bernoulli}(\pi_{ij}),\;\;\pi_{ij}=\frac{\exp(\eta_{ij})}{1+\exp(\eta_{ij})},\\
\eta_{ij} &=& \beta_0+\beta_1x_{ij1}+b_i,\;\;b_i\sim N(0,\sigma^2),\;\;i=1,...,m,j=1,...,n_i,\notag 
\eea
with $\beta_0=-1.5,\ \b_1=2.5,\ \sigma^2=1.5,\ n_i=8$, $x_{ij1}=j/n_i$ and $m=1000$.

\begin{figure}[h]
\centering
\includegraphics[width=1\textwidth,height=.35\textheight]{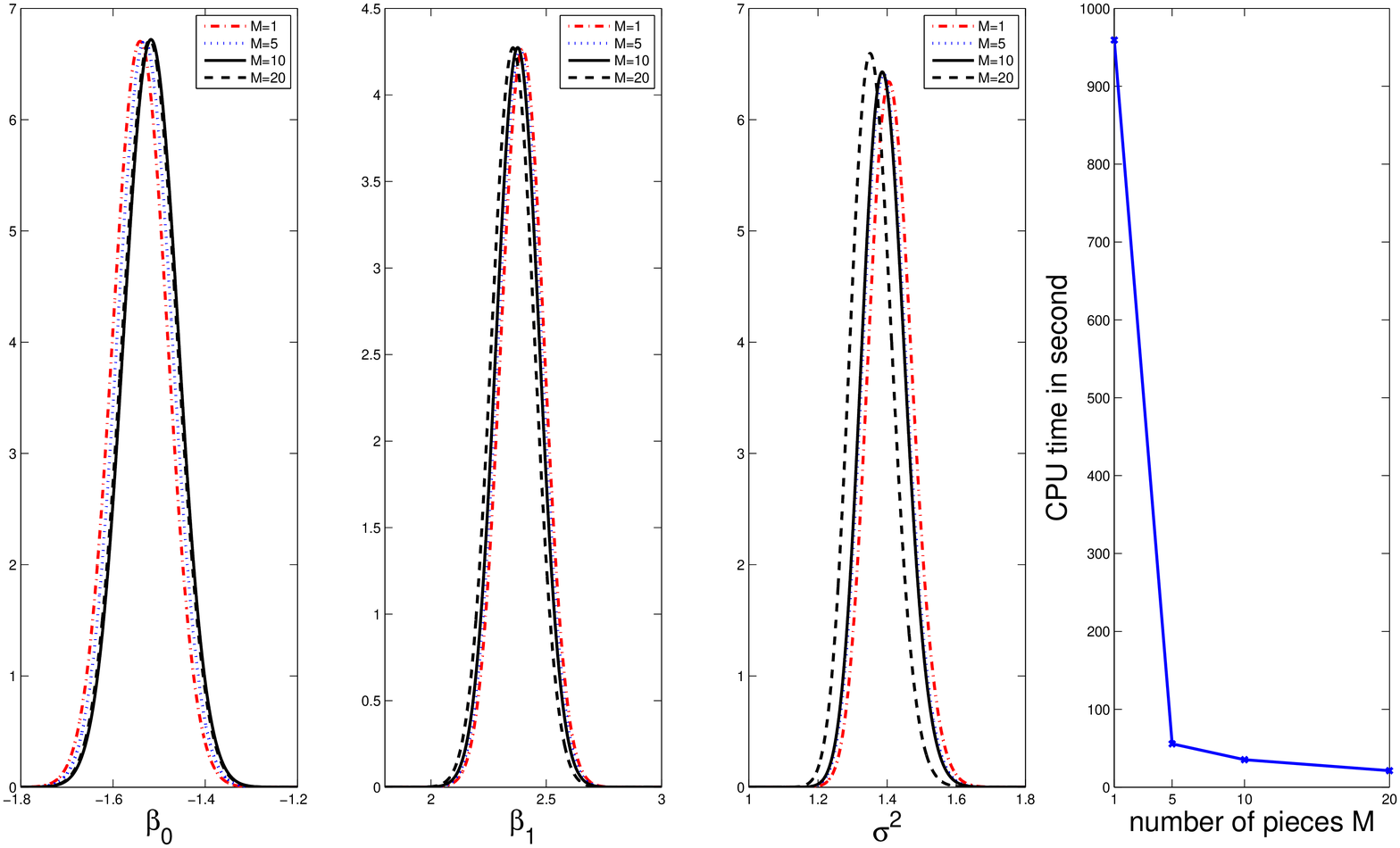}
\caption{Parallel implementation with the four different values of the number of pieces $M$.}
\label{f:parallel_implementation}
\end{figure}

We first generate a dataset from \eqref{e:bino sil}
and run the proposed parallel VB method for four different values of the number of pieces: 
$M=1$ (i.e. no partitions of the data are performed), $M=5$, $M=10$ and $M=20$.
All the partitions are done randomly.
The first three panels of Figure \ref{f:parallel_implementation} plot
the variational densities for $\b_0$, $\b_1$ and $\s^2$ obtained by the four parallel VB runs,
which show that the estimates are close to each other,
in the sense that differences in the estimates are small relative to the estimated posterior standard deviations.
The last panel plots the CPU times taken, which shows that 
running the divide and recombine strategy in parallel gains much efficiency in computing time.   

In order to have a more formal comparison of these four parallel VB runs,
we generate 50 independent datasets from model \eqref{e:bino sil}
and compute the mean squared errors of the estimates of the fixed effects ($\MSE_\b$)
and the mean squared errors of the estimates of the random effect variance ($\MSE_{\s^2}$).
Table \ref{t:parallel simulation} summarizes 
these performance measures and the CPU times averaged over the 50 replications.
The results show that the parallel VB run with $M=5$,
which has 200 subjects on each data piece, produces accurate estimates
while having a reasonable running time.
Our further exploration (results not shown) suggests that we should set $M$ to a value
such that each data piece has roughly 200 subjects in order to have a good tradeoff between computing time and accuracy.  
However, obviously a good choice for the size of each piece depends on factors such as the dimension of the parameter space.  
In practice looking at whether the results change as we divide up the data less finely might be a good data driven diagnostic of whether subset sizes are too small.

All the VB runs in the following examples are run in parallel with $M$ such that each data piece has roughly 200 subjects.
\begin{table}[h]
\centering
\vskip2mm
{\small
\begin{tabular}{ccccc}
\hline\hline
$M$	&$\MSE_\b$	&$\MSE_{\s^2}$	&CPU (second) \\
\hline
1	&0.048		&0.057		&995.9\\
5	&0.048		&0.055		&56.7\\
10	&0.050		&0.061		&36.1\\
20	&0.056		&0.071		&21.6\\
\hline
\hline
\end{tabular}
}
\caption{
The table reports the mean squared errors and the CPU time averaged over 50 replications
for the parallel VB runs with the four values of the number of pieces $M$.
}\label{t:parallel simulation}
\end{table}

\subsubsection{Model selection}
We now study the performance of the model selection procedure discussed in Section \ref{subsec:model selection GLMM}.
We generate datasets from the logistic random intercept model \eqref{e:bino sil}
and also generate covariates $x_{ij2}$ and $z_{ij1}$ randomly from the set $\{-1,0,1\}$.
We have created a model selection problem in which the set of potential covariates
for the fixed effects is $\{1,x_{ij1},x_{ij2}\}$ and for the random effects is $\{1,z_{ij1}\}$.
It is reasonable to always include a fixed intercept and a random intercept in a GLMM,
therefore there are a total of 8 candidate models to consider.
We consider two values of $m$, $500$ and $1000$,
each is used to generate 100 datasets from the true model \eqref{e:bino sil}.
The performance is measured by the correctly fitted rate (CFR) defined as the proportion of
the 100 replications in which the true model is selected.
The CFR is 80\% for $m=500$ and 100\% for $m=1000$,
which shows that the model selection strategy performs well.
The CPU time, averaged over the replications, taken to run the whole model selection procedure 
is 3.54 and 5.86 minutes for $m=500$ and $m=1000$, respectively.
This CPU time is spent on fitting the 8 candidates models and computing the cross-validated LPDS.

\subsubsection{A comparison to MCMC}
This simulation study compares the performance of the proposed parallel and hybrid VB algorithm 
to MCMC. Datasets are generated from a Poisson mixed model with a random intercept
\bean
y_{ij}&\sim&\text{Poisson}(\lambda_{ij}),\;\;\lambda_{ij}=\exp(\eta_{ij}),\\
\eta_{ij} &=& \beta_0+\beta_1x_{ij}+b_i,\;\;b_i\sim N(0,\sigma^2),\;\;i=1,...,m,j=1,...,n_i. 
\eean
We set $\b_0=-1.5,\ \b_1=2.5$, $\s^2=0.2$ and $n_i=5$ with $x_{ij}$ generated from the uniform distribution on $(0,1)$.

The performance is measured by (i) mean squared errors of the estimates of the fixed effects ($\MSE_\b$) and of the estimates of the variance of the random effect ($\MSE_{\s^2}$);
(ii) CPU time in minutes.
Table \ref{t:simulation} reports the simulation result, averaged over 10 replications, for four different sizes of data $m$ ranging from small data ($m=50$) to large data ($m=10000$).
We do not run the MCMC simulation in the case $m=5000$ and $m=10000$ because it is very time consuming.
In the case $m=10000$, it takes approximately 1.1 seconds to run each likelihood estimation in parallel,
thus it would take approximately 733 minutes to run one MCMC chain in the setting of this example.  
Table \ref{t:simulation} shows that the performance of the VB and MCMC is very similar in terms of mean squared errors,
however the VB is much more computationally efficient.  

\begin{table}[h]
\centering
\vskip2mm
{\small
\begin{tabular}{ccccc}
\hline\hline
$m$	&Method		&$\MSE_\b$	&$\MSE_{\s^2}$	&CPU (minute)\\
\hline
50	&VB		&0.155		&0.025		&0.07   \\
	&MCMC		&0.155		&0.057		&18.8\\
200 	&VB		&0.058    	&0.016    	&0.41   \\    
	&MCMC		&0.059		&0.016		&33.4\\
5000	&VB		&0.012		&0.007		&7.4\\
	&MCMC		&- 		&- 		&-\\
10000	&VB		&0.011		&0.004		&14.5\\
	&MCMC		&- 		&- 		&-\\
\hline
\hline
\end{tabular}
}
\caption{
Simulation example. The table reports the mean squared errors and the CPU time averaged over 10 replications.
}\label{t:simulation}
\end{table}

\subsection{Drug longitudinal data}
The anti-epileptic drug longitudinal dataset \citep[see, e.g.,][p.346]{Fitzmaurice:2011} 
 consists of seizures counts on $m=59$ epileptic patients over 5 time-intervals of treatment.
The objective is to study the effects of the anti-epileptic drug on the patients.
Following \cite{Fitzmaurice:2011}, we consider a mixed effects Poisson regression model but with a random intercept
\begin{eqnarray*}
p(y_{ij}|\beta,b_i)&=&\text{Poisson}(\exp(\eta_{ij})),\\
\eta_{ij}&=&c_{ij}+\beta_1+\beta_2\text{time}_{ij}+\beta_3\text{treatment}_{ij}+\beta_4\text{time}_{ij}\times \text{treatment}_{ij}+b_{i},
\end{eqnarray*}
$j=0,1,...,4$, $i=1,...,59$ and $c_{ij}$ is an offset, and $b_i\sim N(0,\s^2)$.
The offset $c_{ij}=\log(8)$ if $j=0$ and $c_{ij}=\log(2)$ for $j>0$,
$\text{time}_{ij}=j$, $\text{treatment}_{ij}=0$ if patient $i$ is in the placebo group
and $\text{treatment}_{ij}=1$ if in the treatment group.   

\begin{figure}[h]
\centering
\includegraphics[width=1\textwidth,height=.15\textheight]{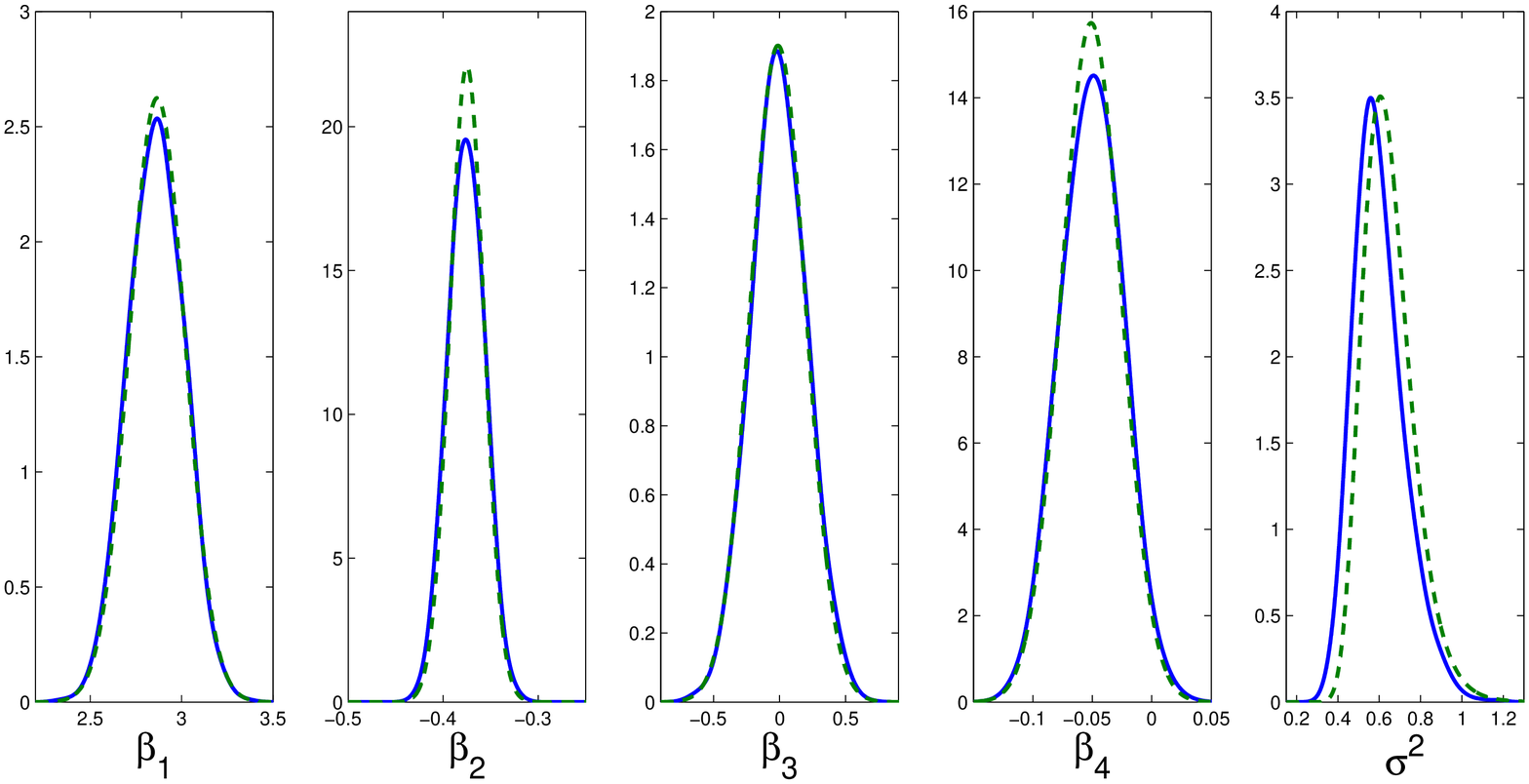}
\caption{The VB estimates (dashed) and MCMC estimates (solid) of the marginal posterior densities for the anti-epileptic drug data.}
\label{f:drug_example}
\end{figure}

The CPU time taken to run the VB and MCMC in this example is 0.14 and 17.7 minutes, respectively.
Figure \ref{f:drug_example} plots the VB estimates (dashed line) and MCMC estimates (solid line) of the marginal posterior densities $p(\b_i|y),\ i=1,...,4$ and $p(\s^2|y)$.   
All the MCMC density estimates in this paper are carried out using the kernel density estimation
based on the built-in Matlab function \texttt{ksdensity}.
The figure shows that the VB estimates are very close to the MCMC estimates in this example.

\subsection{Six city data}
The six cities data in Fitzmaurice and
Laird (1993) consists of binary responses $y_{ij}$ which indicates the wheezing status (1 if wheezing, 0
if not wheezing) of the $i$th child at time-point $j$, $i = 1, . . . , 537$ and $j = 1,...,4$.
Covariates are the age of the child at time-point $j$, centered at 9 years, and the maternal smoking status (0 or 1).
We consider the following logistic regression model with a random intercept
\begin{eqnarray*}
p(y_{ij}|\beta,b_i)&=&\text{Binomial}(1,p_{ij}),\\
\text{logit}(p_{ij})&=&\beta_1+\beta_2\text{Age}_{ij}+\beta_3\text{Smoke}_{ij}+b_{i}.
\end{eqnarray*}

\begin{figure}[h]
\centering
\includegraphics[width=1\textwidth,height=.15\textheight]{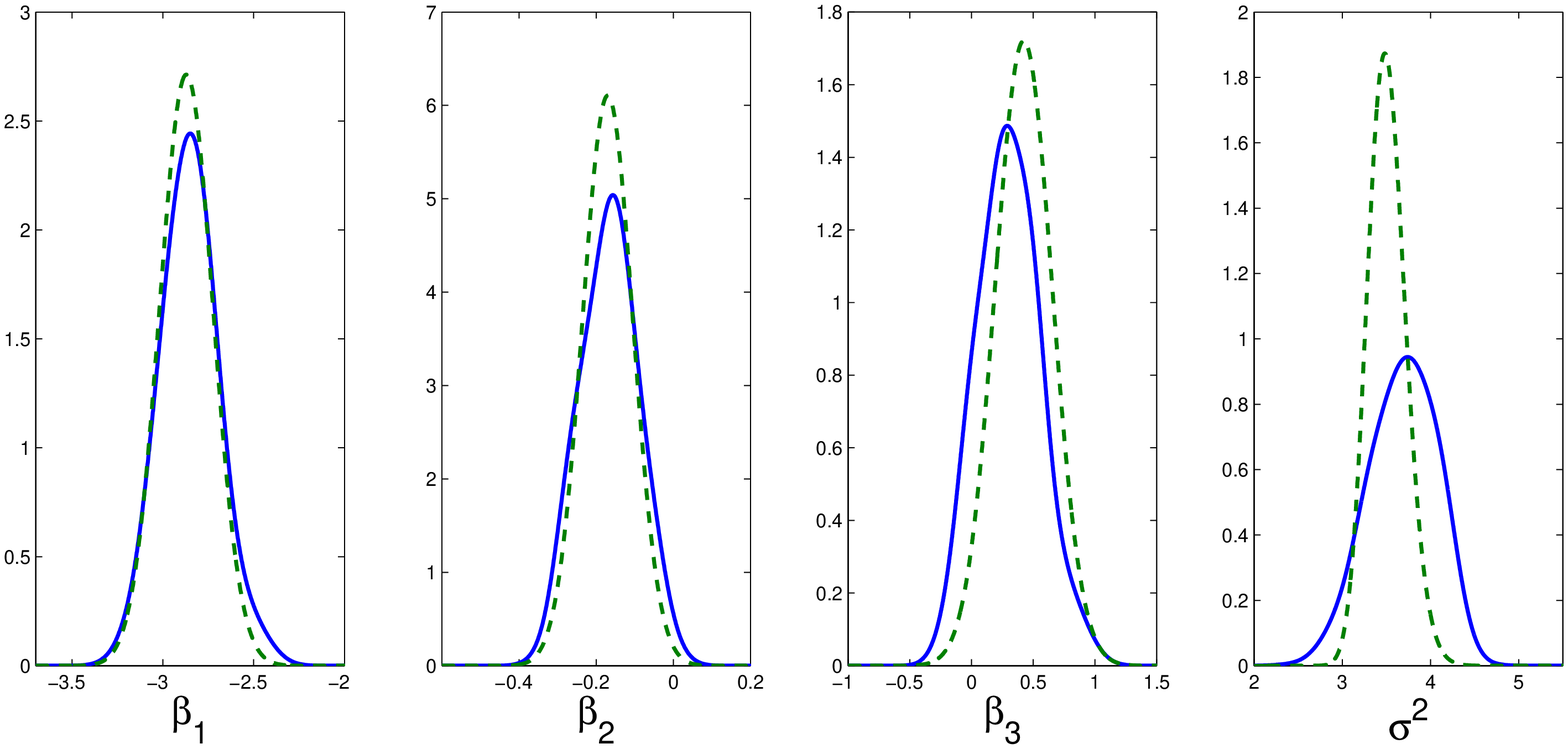}
\caption{The VB estimates (dashed) and MCMC estimates (solid) of the marginal posterior densities for the six city data.}
\label{f:six_city_example}
\end{figure}

Figure \ref{f:six_city_example} plots the VB estimates (dashed line) and MCMC estimates (solid line) of the marginal posterior densities $p(\b_i|y),\ i=1,...,3$ and $p(\s^2|y)$.   
The CPU time taken to run the VB and MCMC in this example is 0.56 and 52.8 minutes, respectively.
The figure shows that the VB estimates of the posterior means are again close to the MCMC estimates.
The VB is about 94 times more computationally efficient than the MCMC implementation considered. 

\subsection{Skin cancer data}
A clinical trial is conducted to test the effectiveness of beta-carotene in preventing
non-melanoma skin cancer \citep{Greenberg:1989}.
Patients were randomly assigned to a control or treatment group
and biopsied once a year to ascertain the number of new skin cancers since the last examination. 
The response $y_{ij}$ is a count of the number of new skin cancers in year $j$ for the $i$th subject. 
Covariates include age, skin (1 if skin has burns and 0 otherwise), gender, exposure (a count of the number of previous
skin cancers), year of follow-up and treatment (1 if 
the subject is in the treatment group and 0 otherwise). 
There are $m = 1683$ subjects with complete covariate information.

\cite{Donohue:2011} consider 5 different Poisson mixed models
with different inclusion of covariates whose including status is given in Table \ref{t:skin example}. 
Using the model selection strategy described in Section \ref{subsec:model selection GLMM},
we compute the cross-validated LPDS whose values are shown in Table \ref{t:skin example},
which suggest that Model 1 should be chosen.
By using an AIC-type model selection criterion,
\cite{Donohue:2011} show that the first three models 
cannot be distinguished and, on parsimony grounds, they select Model 1.

\begin{table}[h]
\centering
\vskip2mm
{\small
\begin{tabular}{l|ccccc}
\hline\hline
			&Model 1&Model 2&Model 3&Model 4&Model 5\\
Fixed intercept		&Y&Y&Y&Y&Y\\
Age 			&Y&Y&Y&Y&Y\\
Skin			&Y&Y&Y&Y&Y\\
Gender			&Y&Y&Y&Y&Y\\
Exposure		&Y&Y&Y&Y&Y\\
Year			&N&Y&Y&Y&Y\\
$\text{Year}^2$		&N&N&Y&N&Y\\
Random intercept	&Y&Y&Y&Y&Y\\
Random slope (Year)	&N&N&N&Y&Y\\
LPDS			&$-277.5$&$-278.5$&$-278.1$&$-1366.6$&$-1404.6$\\
\hline
\hline
\end{tabular}
}
\caption{Five different Poisson mixed models for the skin cancer data and their LPDS values,
which show that Model 1 is chosen.
}\label{t:skin example}
\end{table}

For comparison, after selecting Model 1, we also use MCMC to estimate this model, which is 
\begin{eqnarray*}
p(y_{ij}|\beta,b_i)&=&\text{Poisson}(\exp(\eta_{ij})),\\
\eta_{ij}&=&\beta_0+\beta_1\text{Age}_{i}+\beta_2\text{Skin}_{i}+\beta_3\text{Gender}_{i}+\b_4\text{Exposure}_{ij}+b_{i},
\end{eqnarray*}
where $b_i\sim N(0,\s^2)$, $i = 1,....,m=1683$, $j=1,...,5$.

\begin{figure}[h]
\centering
\includegraphics[width=1\textwidth,height=.15\textheight]{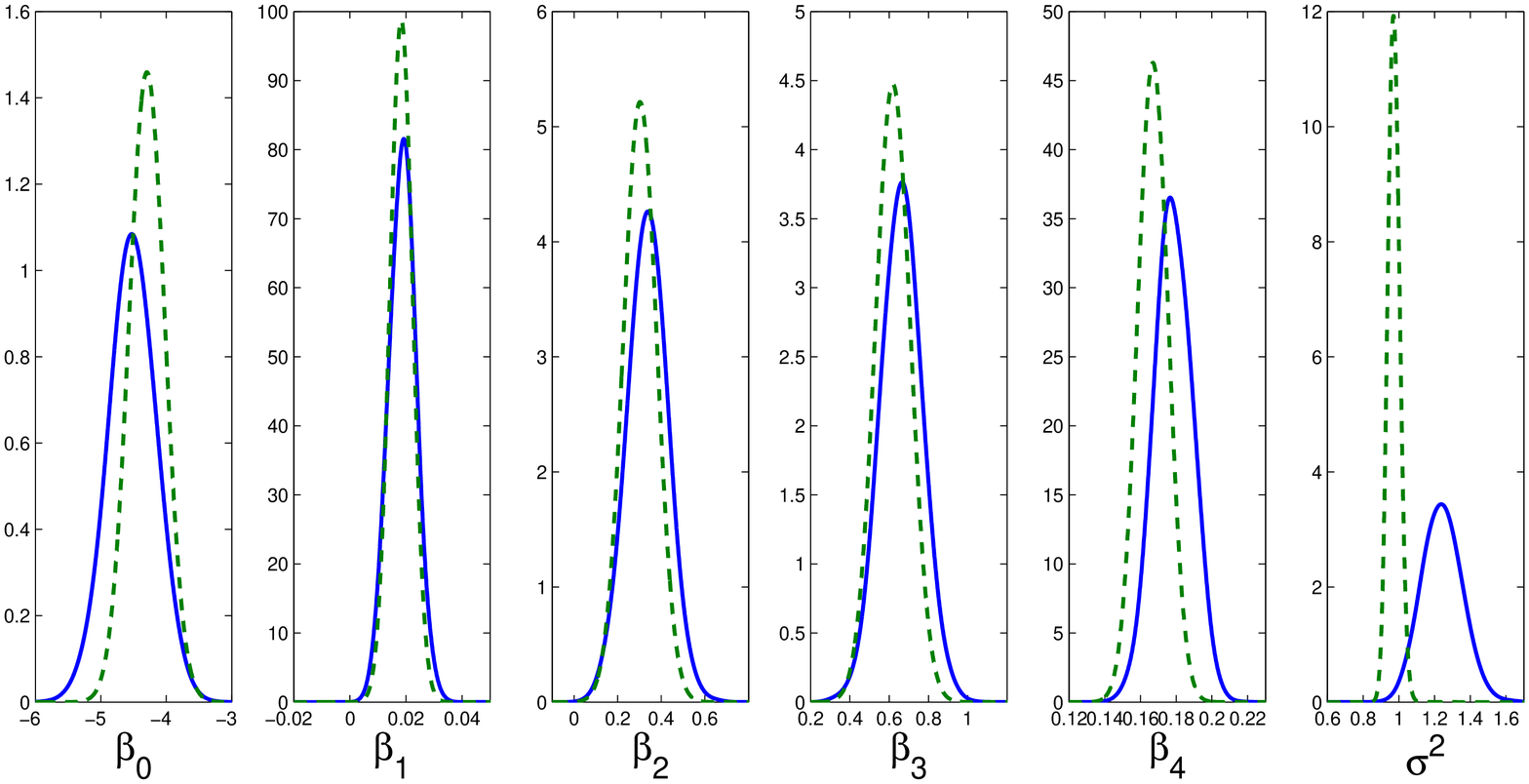}
\caption{The VB estimates (dashed) and MCMC estimates (solid) of the marginal posterior densities when fitting Model 1 to the skin cancer data.}
\label{f:skin_cancer_example}
\end{figure}

Figure \ref{f:skin_cancer_example} plots the VB estimates (dashed line) and MCMC estimates (solid line) of the marginal posterior densities $p(\b_i|y),\ i=0,1,...,4$ and $p(\s^2|y)$.   
The CPU time taken to run the VB and MCMC is 1.45 and 130 minutes, respectively.
The VB and MCMC estimates of the fixed effects $\b$ are pretty similar.
The VB is about 90 times more computationally efficient than the MCMC. 

\section{Conclusion}
We have developed a hybrid VB algorithm that uses a flexible and accurate fixed-form VB algorithm within a mean-field VB updating procedure 
for approximate Bayesian inference,
which is similar in spirit to the Metropolis-Hastings within Gibbs sampling method in MCMC simulation.
If the variational distribution is factorized into a product and 
an exponential form is specified for factors that do not have a conjugate form,
then the new algorithm can be used to approximate any posterior distributions without relying on conjugate priors.
We have also developed a divide and recombine strategy for handling large datasets,
and a method for model selection as a by-product.
The proposed VB method is applied to fitting GLMMs and is demonstrated by several simulated and real data examples.

\section*{Acknowledgment}
The research of Minh-Ngoc Tran and Robert Kohn was partially supported by Australian Research
Council grant DP0667069. 

\bibliographystyle{apalike}
\bibliography{references_v1}

\end{document}